\documentclass[letter]{aa}

\usepackage{amsmath, amssymb, amsfonts, amscd}
\usepackage{mathtools} 
\usepackage{array} 
\usepackage[varg]{txfonts} 

\usepackage{graphicx} 

\usepackage{natbib}
\bibpunct{(}{)}{;}{a}{}{,} 

\usepackage[pdftex=true,pdffitwindow=false,plainpages=false,bookmarksopen=false,breaklinks=true,
pdfpagelabels=true,bookmarksnumbered=true,colorlinks=true, linkcolor=blue]{hyperref} 

\begin{document}

\title{Unravelling tidal dissipation in gaseous giant planets}

\author{
M. Guenel\inst{1}
\and S. Mathis\inst{1,2}
\and F. Remus\inst{3,1}
}

\institute{Laboratoire AIM Paris-Saclay, CEA/DSM - CNRS - Universit\'e Paris Diderot, IRFU/SAp Centre de Saclay, F-91191 Gif-sur-Yvette Cedex, France
\and LESIA, Observatoire de Paris, CNRS UMR 8109, UPMC, Universit\'e Paris-Diderot, 5 place Jules Janssen, 92195 Meudon, France
\and IMCCE, Observatoire de Paris, CNRS UMR 8028, UPMC, USTL, 77 avenue Denfert-Rochereau, 75014 Paris, France\\
\email{mathieu.guenel@cea.fr;stephane.mathis@cea.fr;francoise.remus@obspm.fr} 
}

\date{Received ... / accepted ...}

\abstract
{Tidal dissipation in planetary interiors is one of the key physical mechanisms that drive the evolution of star-planet and planet-moon systems. New constraints on this dissipation are now obtained both in the Solar and exo-planetary systems.}
{Tidal dissipation in planets is intrinsically related to their internal structure. Indeed, the dissipation behaves very differently when we compare its properties in solid and fluid planetary layers. Since planetary interiors consist of both types of regions, it is necessary to be able to assess and compare the respective intensity of the reservoir of dissipation in each type of layers. Therefore, in the case of giant planets, the respective contribution of the potential central dense rocky/icy core and of the deep convective fluid envelope must be computed as a function of the mass and the radius of the core. This will allow to obtain their respective strength.}
{Using a method that evaluates the reservoir of dissipation associated to each region, which is a frequency-average of complex tidal Love numbers, we compare the respective contributions of the central core and of the fluid envelope.}
{In the case of Jupiter and Saturn-like planets, we show that the viscoelastic dissipation in the core could dominate the turbulent friction acting on tidal inertial waves in the envelope. However, the fluid dissipation would not be negligible. This demonstrates that it is necessary to build complete models of tidal dissipation in planetary interiors from their deep interior to their surface without any arbitrary a-priori.}
{We demonstrate how important it is to carefully evaluate the respective strength of each type of dissipation mechanism in planetary interiors and to go beyond the usually adopted ad-hoc models. In the case of gaseous giant planets, we confirm the significance of tidal dissipation in their potential dense core.}


\titlerunning{Unravelling tidal dissipation in gaseous giant planets}
\authorrunning{Guenel, Mathis, Remus}

\maketitle


\section{Introduction and context}

The dissipation of tides is one of the key physical mechanisms that drive the evolution of planetary systems \citep[][]{GS1966}. At the same time, the level of understanding of the related dissipative processes acting both in rocky/icy and in fluid planetary layers remains rather low while they significantly impact the dynamics of star-planet and planet-moon systems \citep[e.g.][]{EL2007,ADLPM2014}. Therefore, a strong effort must be undertaken to get realistic and robust predictions for the rate of dissipation of the kinetic energy of tidal displacements in planetary interiors. In this context, progress are achieved using observational constraints in the Solar and exoplanetary systems \citep[e.g.][]{Laineyetal2009,Husnooetal2012,Albrechtetal2012}. For example, tidal dissipation has been quantified in the cases of Jupiter and Saturn thanks to high-precision astrometric measurements \citep[][respectively]{Laineyetal2009,Laineyetal2012}. These works have demonstrated that these planets are likely to be the seat of a strong dissipation, with in the case of Saturn at least a smooth dependence on the tidal excitation frequency. These results seem to favor the inelastic dissipation in their potential central dense rocky/icy core \citep[e.g.][]{RMZL2012,Storchetal2014}. However, the mass, the size, and the rheology of these cores are still unknown. Moreover, tides excite inertial waves in the deep fluid convective envelope. Their restoring force is the Coriolis acceleration and their dissipation by turbulent friction may be strong and therefore non negligible \citep[e.g.][]{OgilvieLin2004,Ogilvie2013}. As a result, it becomes necessary to explore and to compare the reservoirs of tidal dissipation in each region due to their corresponding dissipative mechanism. This objective must be reached for all kinds of planets since they are potentially all constituted by a combination of solid and fluid layers.

In this first work, we focus on the case of gaseous giant planets. Using simplified two-layer models as an exploratory tool for Jupiter and Saturn-like planets, we apply the method given by \cite{Ogilvie2013} that uses the frequency-dependent Love number to evaluate the reservoirs of dissipation both in their envelope and their core as a function of their mass and aspect ratios. This provides us the first direct evaluation of the relative strength of the different mechanisms of tidal dissipation in a planet, which is constituted by different types of layers. In sec. \ref{sec:modelling}, we describe the main characteristics of our simplified planetary model. Next, we recall the method we use to compute the reservoirs of dissipation due to the viscoelastic dissipation in the core \citep{RMZL2012} and the turbulent dissipation in the fluid envelope \citep{Ogilvie2013}. In sec. \ref{sec:comparison}, we explore their relative strength for realistic values of the radius and the mass of the core and we demonstrate the interest of such an approach. In conclusion, we discuss our results and the potential applications of this method.  
  
\section{Modelling tidal dissipation in gaseous giant planets}
\label{sec:modelling}

\subsection{The studied two-layer model}

\begin{figure}[!t]
\centering
\includegraphics[width=0.3\textwidth]{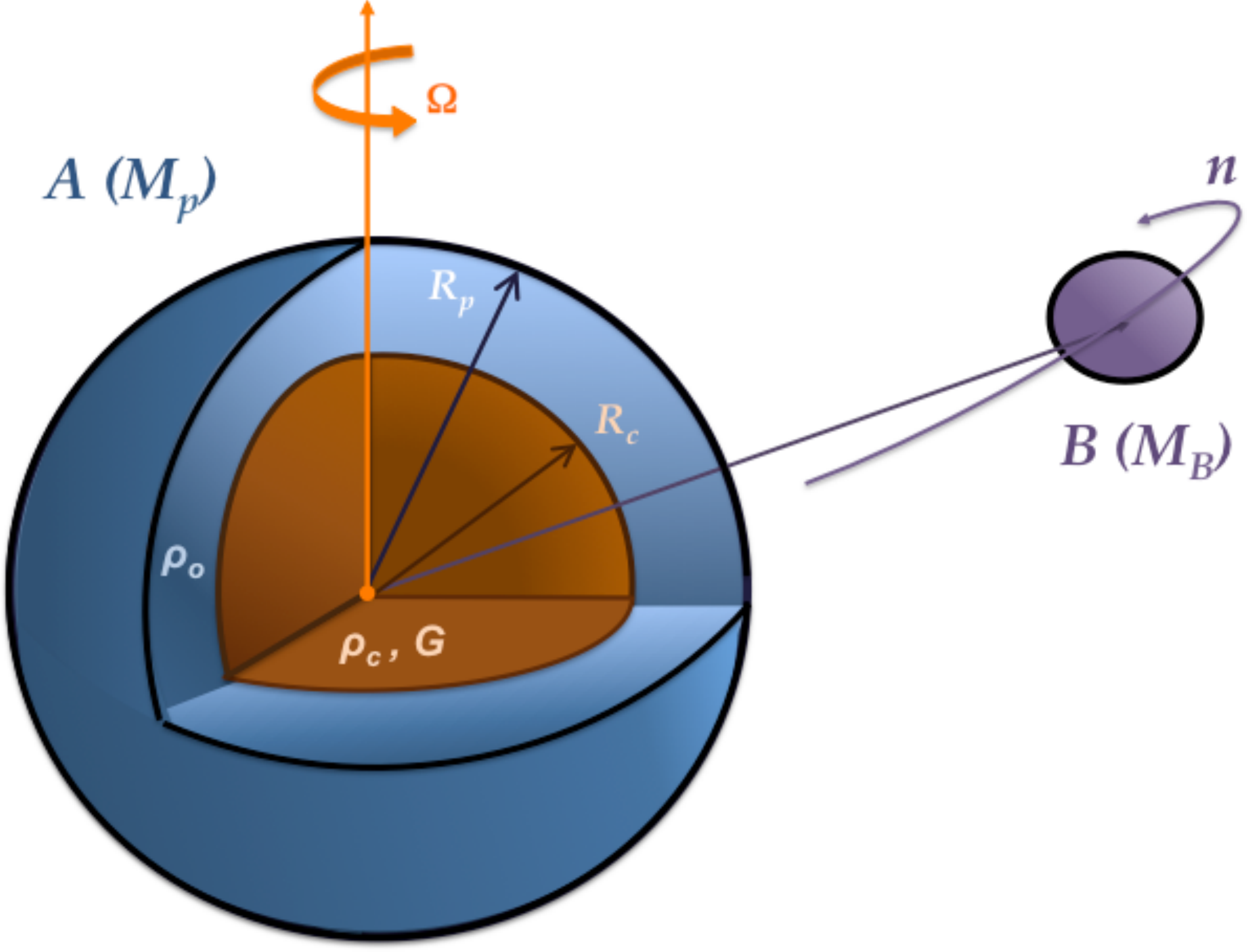}
\caption{Two-layer planet A of mass $M_p$ and mean radius $R_p$ and point-mass tidal perturber B of mass $M_{B}$ orbiting with a mean motion $n$. The rocky/icy solid core of radius $R_c$, density $\rho_c$, and rigidity $G$ (see Eq. \ref{eq:maxwell}) is surrounded by a convective fluid envelope of density $\rho_o$.}
\label{GMR_fig1}
\end{figure}

To study the respective contributions to the tidal dissipation of both the potential rocky/icy core and the fluid envelope of gaseous giant planets, we choose to adopt the simplified two-layer model used in \cite{RMZL2012} and \cite{Ogilvie2013} (see fig. \ref{GMR_fig1}). This model features a central planet A of mass $M_p$ and mean radius $R_p$ along with a point-mass tidal perturber B of mass $m$ orbiting with a mean motion $n$. The body A is assumed to be in moderate solid-body rotation with an angular velocity $\Omega$, so that $\epsilon^2 \equiv \Omega^2/ \sqrt{\mathcal{G} M_p / R_p^3} \ll 1$\footnote{In this regime, the Coriolis acceleration, which scales as $\Omega$, is taken into account while the centrifugal acceleration, which scales as $\Omega^{2}$ is neglected.}, where ${\mathcal G}$ is the gravitational constant. The rocky (or icy) solid core of radius $R_c$ and density $\rho_c$ is surrounded by a convective fluid envelope of density $\rho_o$. Both are assumed to be homogeneous for the sake of simplicity.

\subsection{The evaluation of the tidal dissipation reservoirs}
\label{sec:dissipationreservoir}

The Love numbers quantify the response to the tidal perturbation induced on A by the companion B. More precisely, the Love number $k_l^m$, associated to the $\left(l,m\right)$ component of the time-dependent tidal potential $U$ that corresponds to the spherical harmonic $Y_l^m$, measures at the surface of body A ($r=R_p$) the ratio of the tidal perturbation of its self-gravity potential over the tidal potential. Those numbers are real in the case of perfectly elastic or non-viscous layers and in general they depend on the tidal frequency $\omega=sn-m\Omega$ (with $s\in{\pmb Z}$) \citep[e.g.][]{Efroimsky2012,RMZL2012} just like in any forced oscillating system. However, they turn out to be complex quantities in realistic planetary interiors where dissipation occurs, with a real part that accounts for the energy stored in the tidal perturbation while the imaginary part accounts for the energy losses. Note that ${\rm Im}\left[ k_l^m(\omega) \right]$ is proportional to ${\rm sgn}(\omega)$.

This imaginary part can be expressed in terms of the quality factor $Q_l^m(\omega)$ or equivalently the tidal angle $\delta_l^m(\omega)$, which both depend on the tidal frequency : 
\begin{equation}
{Q_l^m(\omega)}^{-1} = \sin\left[2 \, \delta_l^m(\omega)\right] = {\rm sgn} (\omega)\,{\left| k_l^m(\omega) \right|}^{-1}\,{{\rm Im}\left[ k_l^m(\omega) \right]}.
\end{equation}
Then, following \cite{Ogilvie2013}, we calculate a weighted frequency-average of the imaginary part of the second-order Love number $k_2^2$, which we call the "tidal dissipation reservoir" :
\begin{equation}
\int_{-\infty}^{+\infty} \! {\rm Im} \left[k_2^2(\omega)\right] \,\frac{\mathrm{d}\omega}{\omega} = \int_{-\infty}^{+\infty} \! \frac{\left| k_2^2(\omega) \right|}{Q_{2}^{2}(\omega)} \,\frac{\mathrm{d}\omega}{\omega}.
\label{eq:integralk22}
\end{equation}
This quantity can be defined for any values of $(l,m)$, but we here choose to consider the simplest case of a coplanar system for which the tidal potential ($U$) reduces to the component $(2,2)$  as well as the quadrupolar response of A.

We now examine the two possible mechanisms of dissipation (see fig. \ref{GMR_fig2}): 
\begin{itemize}
\item 
in sec. \ref{sec:inelasticdissipation}, we consider the dissipation associated to the inelasticity of the rocky/icy core following \cite{RMZL2012}; 
\item 
in sec. \ref{sec:turbulentdissipation}, we focus on the dissipation of tidally-excited inertial waves by the turbulent friction in the deep gaseous convective envelope following \cite{Ogilvie2013}. The integral in Eq. \ref{eq:integralk22} then reduces to $\omega \in \left[-2 \Omega,2\Omega\right]$ because higher-frequency acoustic waves are filtered out.
\end{itemize}

\subsection{The inelastic dissipation in the core}
\label{sec:inelasticdissipation}

\begin{figure}[t!]
\centering
\includegraphics[width=0.275\textwidth]{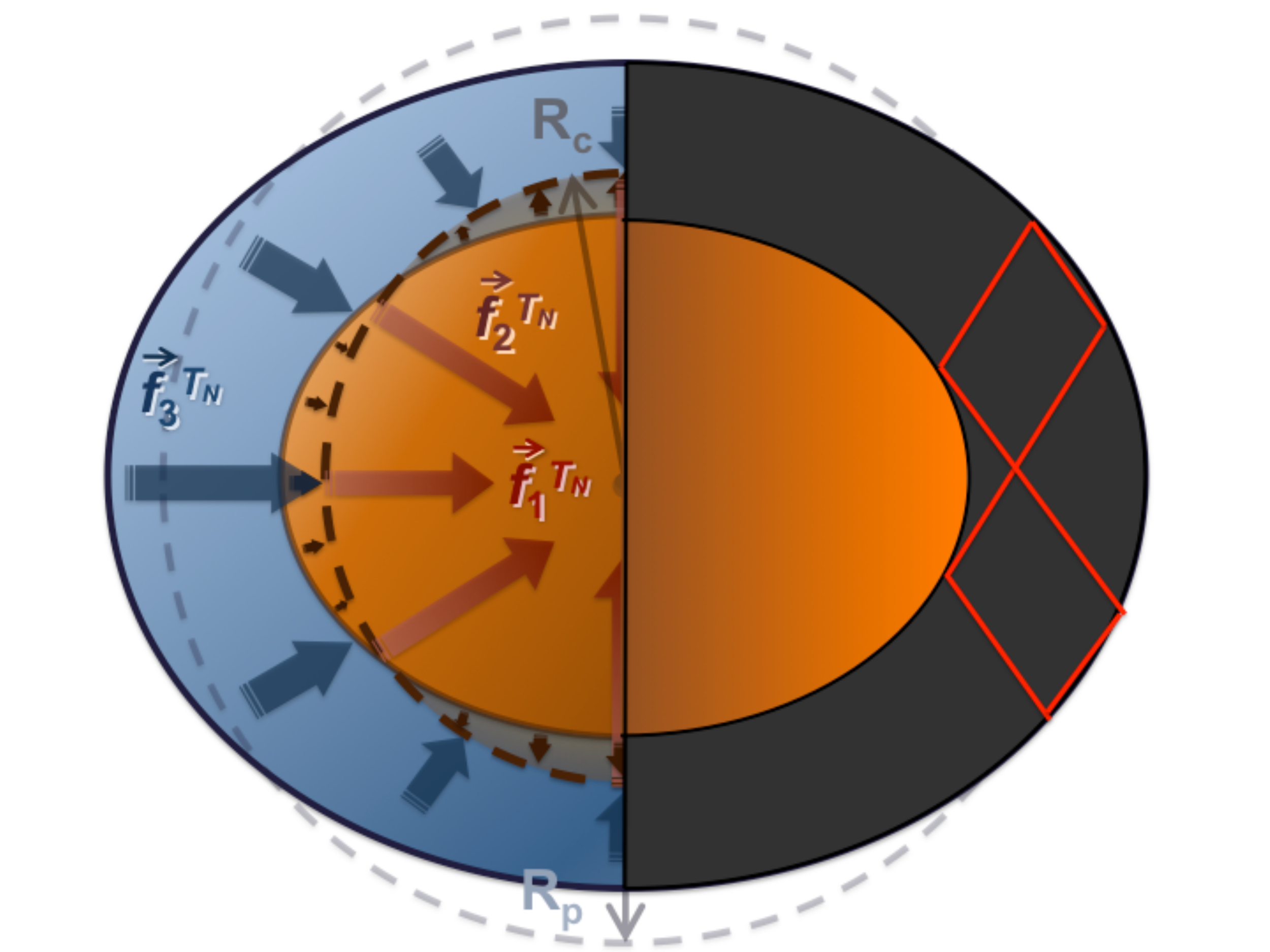}
\caption{Mechanisms of tidal dissipation in our two-layer planetary model: the inelastic dissipation in the dense rocky/icy core (left) and the dissipation due to the tidal inertial waves that reflect onto the core in the fluid convective envelope (right).}
\label{GMR_fig2}
\end{figure}

The inelastic tidal dissipation in the solid core is the result of its internal viscosity ($\eta$). In the case of the studied two-layer model, it is modified by the set of mechanical constraints, namely the gravitational forces (${\vec f}_{1}^{T_N}$), the loading of the core due to its deformation (${\vec f}_{2}^{T_N}$), and the hydrostatic pressure exerted by the surrounding fluid envelope (${\vec f}_{3}^{T_N}$), which is here assumed to be static and non dissipative \citep[see][for a complete discussion and fig. \ref{GMR_fig2}]{Dermott1979,RMZL2012}. 

Following \cite{RMZL2012} and \cite{Remus2013}, the second-order Love number $k_2^2(\omega)$ is given by:
\begin{equation}
\label{eq:k22remus}
k_2^2(\omega) = \frac{\tilde{H}+\alpha+3}{\frac{2}{3}\alpha\tilde{H}-\frac{3}{2}},
\end{equation}
where $\alpha$ and $\tilde{H}$ are functions of the aspect ratio $(R_c/R_p)$, the density ratio $(\rho_o/\rho_c)$, and the complex effective shear modulus $\hat{\mu}$ of the core : 
\begin{align*}
&\alpha = 1 + \frac{5}{2}\frac{\rho_c}{\rho_o} \left( \frac{R_c}{R_p} \right)^{3} \left( 1- \frac{\rho_o}{\rho_c} \right), \\
&\tilde{H} = \beta \left[ \left( 1+ \frac{3}{2} \frac{\rho_o}{\rho_c} \right) \left( 1- \frac{\rho_o}{\rho_c}\right) + \hat{\mu}(\omega) \right], \\
&\beta = \left( \frac{R_c}{R_p} \right)^{-5} \left( 1- \frac{\rho_o}{\rho_c} \right)^{-2}, \quad \frac{\hat{\mu}(\omega)}{\bar{\mu}(\omega)} = \gamma = \frac{19}{2 \, \rho_c \, g_c \, R_c},
\end{align*}
where $\bar{\mu}$ is the complex shear modulus and $g_c$ is the gravity at $r=R_c$. Note also that ${\rm Im} \left[k_2^2(\omega)\right] $ scales as $( R_c/R_p )^5$ as $R_c/R_p \rightarrow 0$. This result is valid for any linear rheology but the mechanical behavior of the dense central rocky/icy cores in gaseous giant planets is poorly constrained \citep[see e.g.][]{Henningetal2009}. For that reason, we use the simplest linear viscoelastic Maxwell model for which
\begin{equation}
\label{eq:maxwell}
{\rm Re}\left[\bar{\mu}(\omega)\right] = \frac{\eta^2 \, G \, \omega^2}{G^2+\eta^2 \, \omega^2}\quad\hbox{and}\quad {\rm Im}\left[\bar{\mu}(\omega)\right]= \frac{\eta \, G^2 \, \omega}{G^2+\eta^2 \, \omega^2} ,
\end{equation}
where $G$ is the rigidity and $\eta$ is the viscosity \citep[see][]{Henningetal2009,RMZL2012}. For this model, the core behaves as a rigid body when $\omega\gg\omega_{M}$, and as a fluid body when $\omega\ll\omega_{M}$, where $\omega_{M}=G/\eta$ is the Maxwell frequency. We find that
\begin{align}
&\int^{+\infty}_{-\infty} \! {\rm Im} \left[k_2^2(\omega)\right] \,\frac{\mathrm{d}\omega}{\omega}  = \frac{\pi \,G \left(3 + 2 \alpha\right)^2 \beta\, \gamma}{\delta \left(6\,\delta+4\,\alpha\,\beta\,\gamma\, G\right)},\label{viscoelastic_reservoir}\\[5pt]
&\hbox{with}\quad\delta = \left[\frac{2}{3} \,\alpha\, \beta\, \left(1 - \frac{\rho_o}{\rho_c}\right) \left(1 + \frac{3}{2} \frac{\rho_o}{\rho_c}\right) - \frac{3}{2}\right],\nonumber
\end{align}
which is remarkably independent on the viscosity $\eta$ and vanishes for small values of $G$.

\subsection{The dissipation of inertial waves in the envelope}
\label{sec:turbulentdissipation}

Tidal dissipation in the fluid convective envelope of A originates from the excitation by B of inertial waves, which are driven by the Coriolis acceleration. They are damped by the turbulent friction, which can be modeled using a turbulent viscosity \citep{OL2012}. Its evaluation in our two-layer model was conducted by \cite{Ogilvie2013} who assumed an homogeneous and perfectly rigid solid core where no inelastic dissipation occurs, while the envelope is homogenous and incompressible. The solutions of the system of dynamical equations for the fluid envelope written in the co-rotating frame are separated into a non-wavelike part (with subscripts $_{\rm nw}$), which corresponds to the immediate hydrostatic adjustment to the external tidal potential ($U$), and a wavelike part (with subscript $_{\rm w}$) driven by the action of the Coriolis acceleration on the non-wavelike part :
\begin{equation}
\begin{cases}
\ddot{\mathbf{s}}_{\mathrm{nw}} = -\nabla W_{\mathrm{nw}} ,\\
h'_{\mathrm{nw}} + \Phi'_{\mathrm{nw}} + U = 0 ,\\
\rho_{\mathrm{nw}}' = -\nabla \cdot (\rho \, \mathbf{s}_{\mathrm{nw}}), \\
\nabla^2 \Phi_{\mathrm{nw}}' = 4 \, \pi \, \mathcal{G} \, \rho_{\mathrm{nw}}',
\end{cases}
\hbox{and}\quad
\begin{cases}
\ddot{\mathbf{s}}_{\mathrm{w}} + 2 \, \Omega \, \mathbf{e}_z \times \dot{\mathbf{s}}_{\mathrm{w}} = -\nabla W_{\mathrm{w}} + \mathbf{f} ,\\
h'_{\mathrm{w}} = \Phi'_{\mathrm{w}} = \rho_{\mathrm{w}}' = 0, \\
\nabla \cdot (\rho \, \mathbf{s}_{\mathrm{w}}) = 0, 
\end{cases}
\end{equation}
where $\mathbf{s}$ is the displacement, $\mathbf{e}_z$ the unit vector along the rotation axis, $h$ the specific enthalpy, $\Phi$ the self-gravitational potential of A, and $\rho$ is the density. Primed variables denote an Eulerian perturbation in relation to the unperturbed state with unprimed variables. Note that $U$ and $\mathbf{s}$ are actually perturbations too. Finally, $W \equiv W_{\mathrm{nw}}+ W_\mathrm{w} = h' + \Phi' + U$ while $\mathbf{f} = -2 \, \Omega \, \mathbf{e}_z \times \dot{\mathbf{s}}_{\mathrm{nw}}$ is the acceleration driving the wavelike part of the solution.

The kinetic energy of the wavelike part of the solution can be derived without solving the whole system of equations, thanks to an impulsive calculation. This kinetic energy will eventually be dissipated (no matter the exact physical processes at stake here, even if we know that it can be modeled by a turbulent viscosity) and is related to the tidal dissipation reservoir introduced in Eq. (\ref{eq:integralk22}). The final result is \citep{Ogilvie2013} :
\begin{eqnarray}
\lefteqn{\int^{+\infty}_{-\infty} \! {\rm Im} \left[k_2^2(\omega)\right] \,\frac{\mathrm{d}\omega}{\omega} = \frac{100 \pi}{63} \epsilon^2 \displaystyle{\frac{\left( R_c/R_p \right)^5}{1-\left( R_c/R_p \right)^5}}}\label{eq:imk22ogilvie}\\
&&\times\left[ 1+ \frac{1-\rho_o / \rho_c}{\rho_o / \rho_c} \left( R_c/R_p \right)^3 \right]\left[ 1+ \frac{5}{2} \frac{1-\rho_o / \rho_c}{\rho_o / \rho_c} \left( R_c/R_p \right)^3 \right]^{-2}\!.\nonumber
\end{eqnarray}

\section{Comparison of the two dissipation mechanisms}
\label{sec:comparison}

\begin{figure}[t!]
\centering
\includegraphics[width=0.35\textwidth]{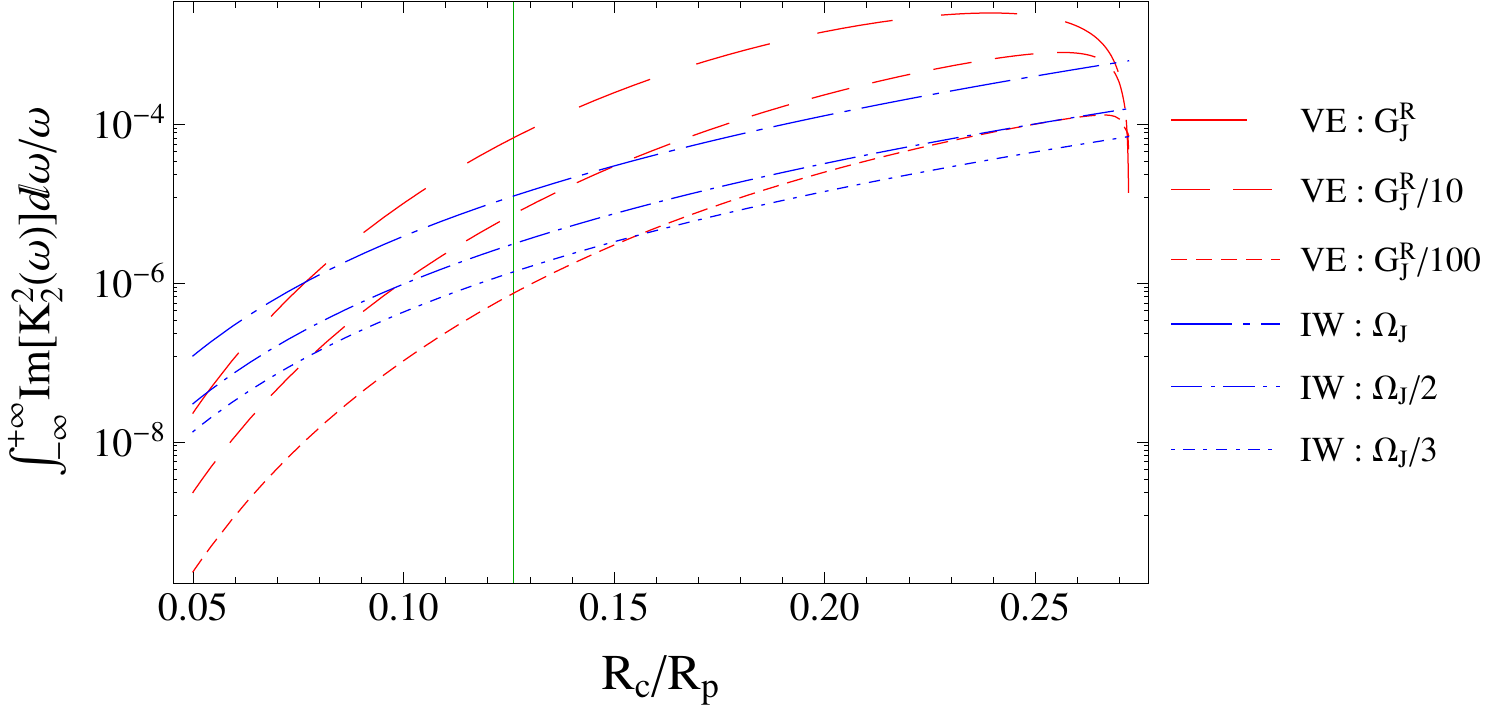}
\includegraphics[width=0.35\textwidth]{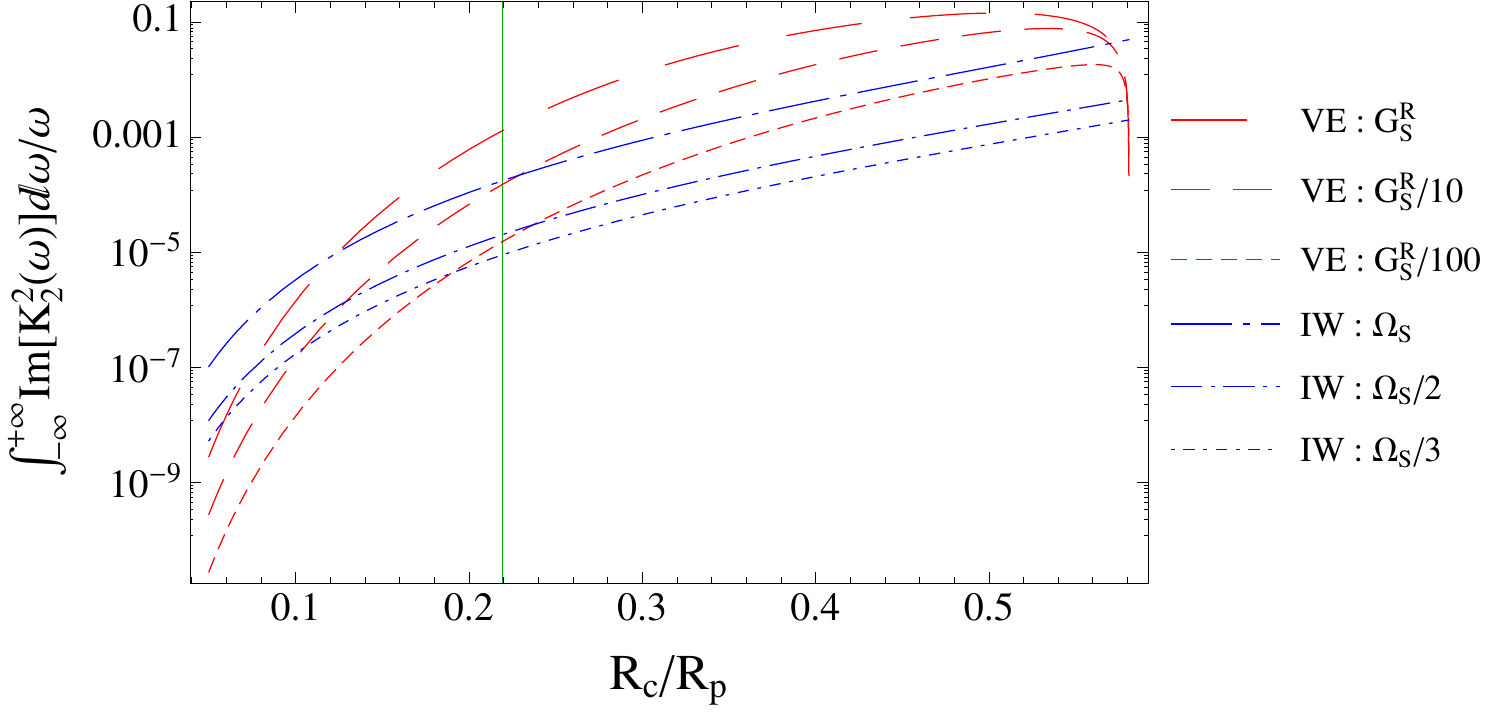}
\caption{Dissipation reservoirs for the viscoelastic dissipation in the core (red curve) and the turbulent friction in the fluid envelope (blue curves) in Jupiter- (above) and Saturn-like planets (below) as a function of the aspect ratio $R_c/R_p$, the rotation rate $\Omega$, and the rigidity of the core $G$, with fixed $R_p$ and $M_p$. We use the values $M_c/M_p = \left\{0.02,0.196\right\}$ from \cite{Guillot1999} and \cite{Hubbardetal2009} for Jupiter and Saturn respectively. The vertical green line corresponds to $R_c/R_p = \left\{0.126,0.219\right\}$.}
\label{GMR_fig3}
\end{figure}

\begin{figure}[t!]
\centering
\includegraphics[width=0.375\textwidth]{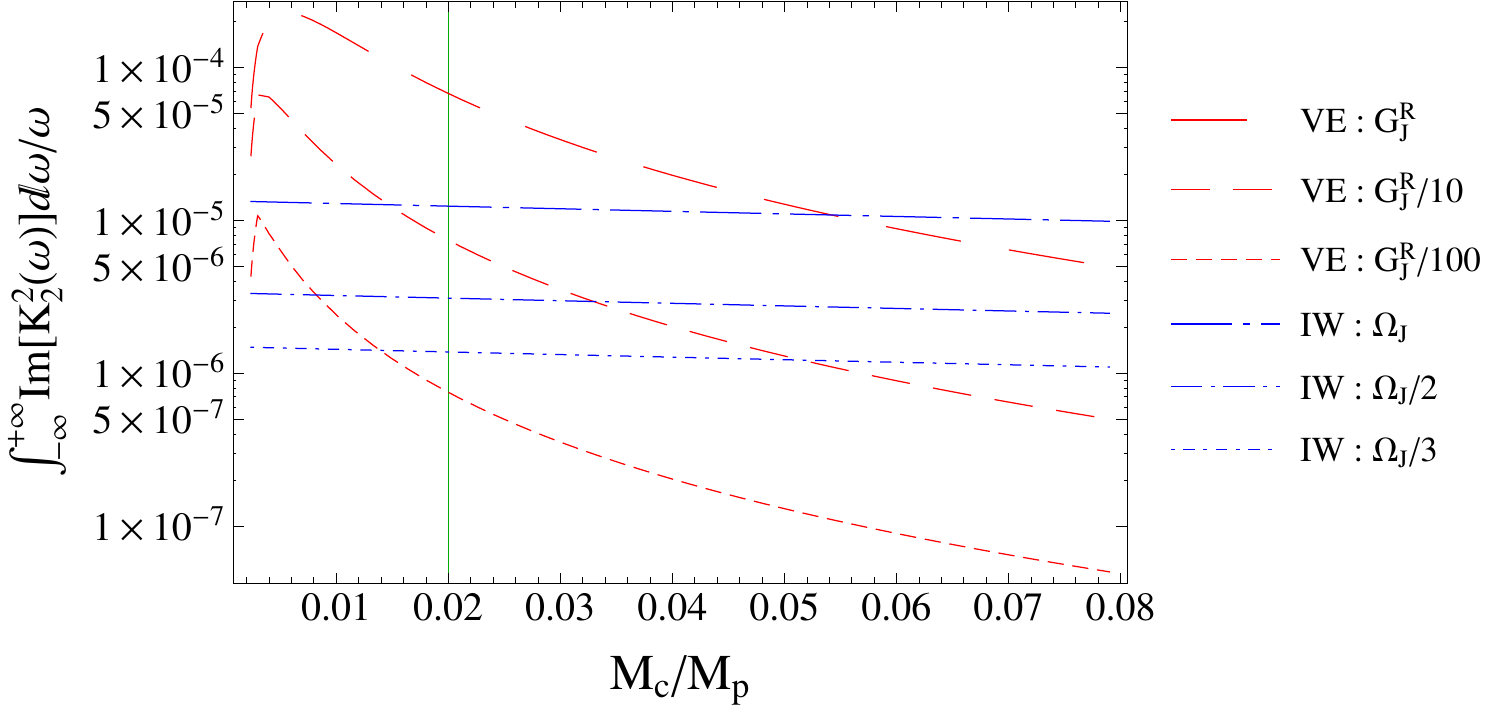}
\includegraphics[width=0.375\textwidth]{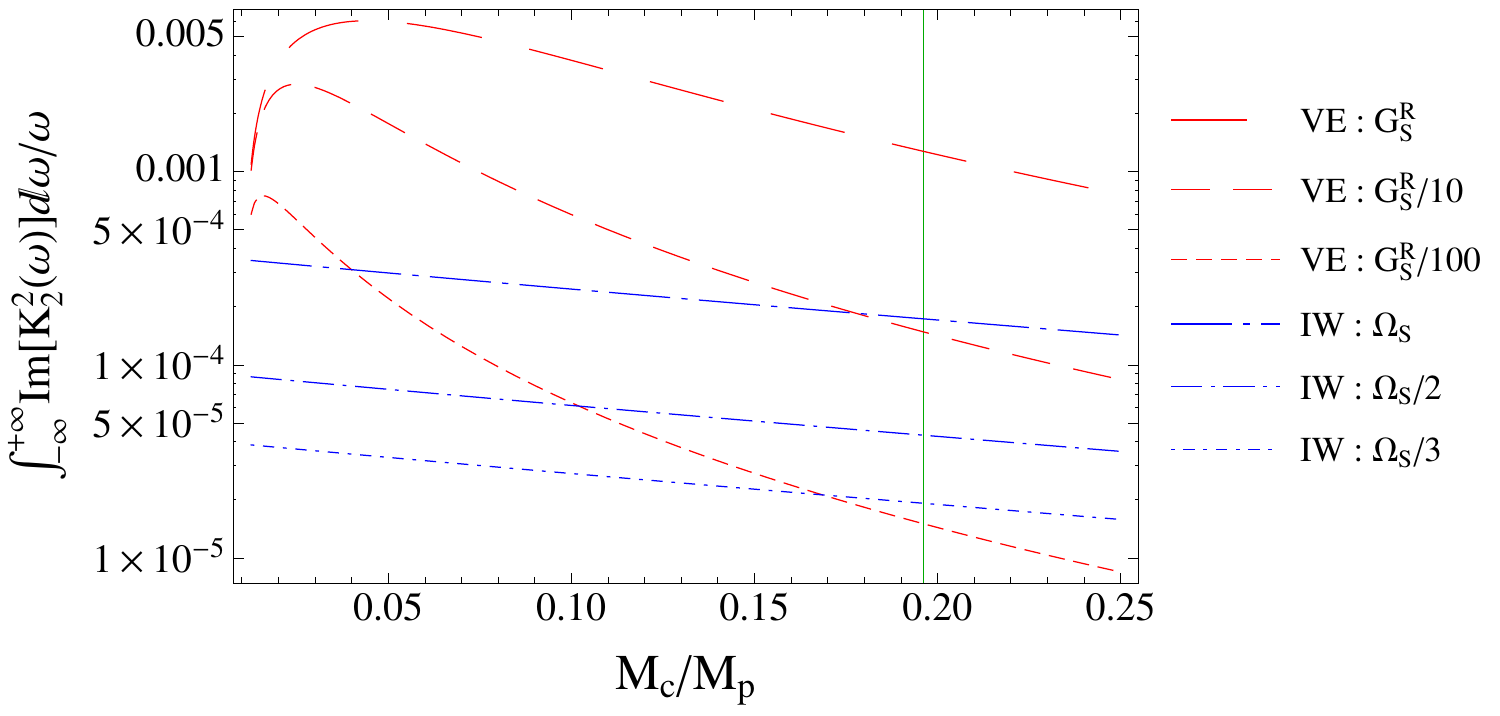}
\caption{Same as fig. \ref{GMR_fig3} but as a function of the mass ratio $M_c/M_p$ with fixed $M_p$ and $R_p$. We adopt $R_c/R_p = \left\{0.126,0.219\right\}$ for Jupiter and Saturn respectively. The wide $M_c$-ranges [1,3 - 25] $M_{\oplus}$ for Jupiter and [2 - 24] $M_{\oplus}$ for Saturn are those considered as possible by \cite{Guillot1999,Nettelmann2011,Nettelmann2013}. The vertical green line corresponds to $M_c/M_p = \left\{0.02,0.196\right\}$.}
\label{GMR_fig4}
\end{figure}

Our goal is to compare quantitatively the respective strength of the two dissipation mechanisms in order to determine if and when either one of them can be neglected in gaseous giant planets similar to Jupiter and Saturn. Their respective mass and radius are $M_p=\left\{317.83,95.16\right\}M_{\oplus}$ and $R_p=\left\{10.97,9.14\right\}R_{\oplus}$ with $M_{\oplus}=5.97\,10^{24}$ kg and $R_{\oplus}=6.37\,10^3$ km being the Earth's mass and radius. Their rotation rate are $\Omega_{\left\{{\rm J,S}\right\}}=\left\{1.76\,10^{-4},1.63\,10^{-4}\right\}{\rm s}^{-1}$. Internal structure models for these bodies are still not well constrained. This is why we choose to explore wide ranges of core radii in fig. \ref{GMR_fig3} (covering the values considered possible by \cite{Guillot1999} for Jupiter and \cite{Hubbardetal2009} for Saturn) and core masses in fig. \ref{GMR_fig4} (covering the values considered possible by \cite{Guillot1999}, \cite{Nettelmann2011}, and \cite {Nettelmann2013}). In order to do this, we need to use fixed values for the mass ratios $M_c/M_p$ (in fig. \ref{GMR_fig3}) or for the aspect ratios $R_c/R_p$ (in fig. \ref{GMR_fig4}), along with specific values of the angular velocity, $\Omega$, for tidal inertial waves (eq. \ref{eq:imk22ogilvie}) and of the rigidity $G$ (eq. \ref{viscoelastic_reservoir}) for the viscoelastic model. We choose to use as a reference $G_{\left\{{\rm J,S}\right\}}^{\rm R}=\left\{4.46\,10^{10},1.49\,10^{11}\right\}\,{\rm Pa}$ that allow this dissipation model to match the dissipation measured by \cite{Laineyetal2009, Laineyetal2012} in Jupiter at the tidal frequency of Io and in Saturn at the frequency of Enceladus (with $\eta_{\left\{{\rm J,S}\right\}}=\left\{1.45\,10^{14},5.57\,10^{14}\right\}\,{\rm Pa}\cdot{\rm s}$). We assume the core masses proposed by \cite{Guillot1999} and \cite{Hubbardetal2009}, i.e. $M_c=\left\{6.41,18.65\right\}M_{\oplus}$ that yields the minimum core's radii $R_c=\left\{0.126,0.219\right\}R_p$. This allows us to avoid any underestimation of the solid dissipation reservoir that a poor choice of parameters could cause \citep[see fig. 9 in][]{RMZL2012}.

\subsection{As a function of the core radius}

Figure \ref{GMR_fig3} shows that for both dissipation models and both planets, the tidal dissipation reservoirs generally increase with the core radius until a critical value is reached, where $\rho_o/\rho_c = 1$, which is a singularity of the model, the density ratio decreasing with the core radius since $R_p$ and $M_c/M_p$ are fixed. Here, we adopt the values $M_c/M_p = \left\{0.02,0.196\right\}$ given by \cite{Guillot1999} and \cite{Hubbardetal2009} respectively for Jupiter and Saturn. This result is in agreement with the predictions of \cite{RMZL2012} for the core and of \cite{OgilvieLin2004}, \cite{GoodmanLackner2009}, \cite{RV2010}, and \cite{Ogilvie2013} who explain that inertial waves in a fluid spherical shell experience multiple reflections on its boundaries and follow specific paths called attractors, where shear layers take place, leading to an enhanced viscous dissipation compared to the case of a full sphere \citep[][and fig. \ref{GMR_fig2}]{Wu2005}. These plots show that in Jupiter- and Saturn-like gaseous giant planets, the two distinct mechanisms exposed in sec. \ref{sec:modelling} can both contribute to tidal dissipation, and that therefore none of them can be neglected in general. Moreover, when $R_c/R_p>\left\{0.126,0.219\right\}$, $\Omega=\Omega_{\left\{{\rm J,S}\right\}}$, and $G=G_{\left\{{\rm J,S}\right\}}^{\rm R}$, the viscoelastic dissipation slightly dominates the one in the fluid envelope until the singularity of the model is reached.

\subsection{As a function of the core mass}

If we now study the problem as a function of the core mass, we observe that the two tidal dissipation reservoirs associated to each model slightly decrease because the density ratio $\rho_o/\rho_c$ decreases since $M_p$ and $R_c/R_p$ are fixed. Here, we adopt $R_c/R_p=\left\{0.126,0.219\right\}$ for Jupiter and Saturn respectively following previous sections. Again, the order of magnitude of each dissipation reservoir can be similar and there is no clear indication that either one of them is negligible. In the case where $M_c/M_p=\left\{0.02,0.196\right\}$, $\Omega=\Omega_{\left\{{\rm J,S}\right\}}$, and $G=G_{\left\{{\rm J,S}\right\}}^{\rm R}$, the viscoelastic dissipation slightly dominates the one in the fluid envelope.

\section{Conclusions and perspectives}

In this work, we computed for the first time a direct comparison of the relative strength of tidal dissipation mechanisms in the interiors of gaseous giant planets. Even if it is necessary to keep in mind that this quantitative comparison is obtained using simplified two-layer planetary models, we are confident that this approach is robust enough to explore and to evaluate the amplitude of both solid and fluid tidal dissipations and to compare them. In this framework, we find that to be able to reproduce the observed values of the tidal dissipation in Jupiter and in Saturn obtained thanks to high-precision astrometry \citep{Laineyetal2009,Laineyetal2012}, we are in a situation where the viscoelastic dissipation in the core may dominate the turbulent friction acting on tidal inertial waves in the envelope. However, the fluid mechanism is not negligible which demonstrates the necessity to compute models that take into account all the possible dissipation mechanisms for complex planetary interiors. The action of each of them on the spins of bodies constituting planetary systems and on their orbital architecture would be unravelled thanks to their behaviour as a function of the excitation frequency \citep{ADLPM2014} and of realistic formation/evolution simulations \citep[e.g.][]{Charnozetal2011,Laskaretal2012}. Moreover, this method that uses frequency-dependent complex Love numbers would be applied in a near future to the case of realistic stratified solid and fluid regions \citep[e.g][]{OgilvieLin2004,Tobieetal2005} and to other types of planets such as icy giant planets and super-Earths, which are also composed by a superposition of both solid and fluid regions.

To get robust predictions from ab-initio treatment of the mechanisms of tidal dissipation, it would be also necessary to improve simultaneously our understanding of the rheological behaviour of rocky and icy planetary layers and to take into account possible stable stratification, differential rotation, magnetic fields, and non-linear processes such as instabilities and turbulence in fluid regions. 

\begin{acknowledgements}
We thank the referee who allowed to improve the manuscript. M. Guenel was supported by CEA. This work was funded partly by the Programme National de Plan\'etologie (CNRS/INSU), the Campus Spatial de l'Universit\'e Paris Diderot, the "L'Or\'eal-Acad\'emie des Sciences-Unesco" fundation, and the Emergence-UPMC project EME0911 Encelade.
\end{acknowledgements}

\bibliographystyle{aa}  
\bibliography{GMRref} 

\end{document}